# Cool, But What About Oracles? – An Oracle-Based Perspective on Blockchain Integration in the Accounting Field.

Giulio Caldarelli
University of Turin (Italy).
giulio.caldarelli@unito.it
ORCID-ID: 0000-0002-8922-7871

**ABSTRACT**

The Bitcoin Network is a sophisticated accounting system that allows its underlying cryptocurrency to be trusted even in the absence of a reliable financial authority. Given its undeniable success, the technology, generally referred to as blockchain, has also been proposed as a means to improve legacy accounting systems. Accounting for real-world data, however, requires the intervention of a third party known as an Oracle, which, having not the same characteristics as a blockchain, could potentially reduce the expected integration benefit. Through a systematic review of the literature, this study aims to investigate whether the papers concerning blockchain integration in accounting consider and address the limitations posed by oracles. Results support the view that although research on the subject counts numerous articles, actual studies considering oracle limitations are lacking. Interestingly, despite the scarce production of papers addressing oracles in various accounting sectors, reporting for ESG already shows interesting workarounds for oracle limitations, with permissioned chains envisioned as a valid support for the safe storage of sustainability data.

## Introduction

Defined as a "Truth Machine," [1] blockchain has been widely advised for applications and purposes in which the lack of trust required some additional trust mechanisms. However, due to inherent technical limitations and the reliance on external data providers known as oracles, blockchain is unable to guarantee the authenticity or reliability of data stored, and therefore, optimal use cases are to be found in contexts in which the trusted reliability of data is not questioned [2]. On the other hand, If leveraged to guarantee the inalterability and provenance of data, it is certainly helpful when properly implemented. For example, digital signatures, proof of work, hashing, and time-stamps of the Bitcoin network can be efficiently leveraged to authenticate academic records, guaranteeing they are issued from the related academic institution [3], [4]. In the field of legacy accounting, blockchain has been explored as a possible solution to improve the reliability of accounting information, deter against concealment, provide proof of transactions, prevent data loss and manipulation, ensure traceability of transactions, and finally improve control over management actions [5], [6], [7], [8]. Above a general improvement of accounting data and practices, blockchain has also been envisioned as a possible enabler for specific technological innovations such as real-time accounting, whose intent is to dramatically reduce the time gap between transaction registration and verification [9]. This time-discrepancy reduction is sought to limit and, where possible, also prevent malpractice, such as earning management [10], [11]. Given the blockchain's characteristics of transparency and immutability, a stream of literature also hypothesizes its use as an enabler for triple-entry accounting, an old theoretical construct, revamped with the work of Grigg [12] aimed at reducing accounting discrepancies within the same transactions registered by different entities. A further stream of research envisioned the potential of blockchain as a means to enhance ESG reporting by improving the reliability of sustainability accounting data [13], [14]. Depending on the perception of the potential of blockchain technology, different integration types are proposed and for different finalities. When properly implemented and when needed blockchain can undeniably support the related applications and sectors. However, If implemented for uses that are outside its scope, due to the Oracle problem, blockchain is just a really expensive database made of "immutable garbage [15]". The case described before of the academic institution is very explicative of the oracle problem. Blockchain in fact ensures that the certificate is authentic and has been signed by the institution but does not affect the reliability over its content. If the institution is not trusted or is known for emitting false certificates, the fact that emitted certificates are on a blockchain does not improve their reliability or truthfulness in any way [4]. In the case of accounting, the situation is definitely more complex since a massive number of transactions result in a large quantity of data. Moreover, we have more actors with different incentives to provide true or false data and different stakeholders on which recorded data may have externalities. Therefore, oracles intended as the mechanism by which accounting data is securely gathered and transmitted to the blockchain are fundamental for its reliability.

Therefore, this article's scope is to review the academic literature published on blockchain integration in the accounting field and inspect if the limitations of oracles and the oracle problem have been efficiently considered and addressed. Data will be extracted building on a previous review with a similar purpose [16], but focused on accounting literature and enriched with additional variables. As previous research supports the idea that by overlooking oracles and oracle problem studies may include misconceptions about the potential of blockchain, the present research will also investigate the presence within the published research of common misconceptions on blockchain potential. The idea is to enlighten the rationale for implementing blockchain in the accounting field and what advantage is sought to be obtained to determine if the expected outcome is within the boundaries of technology potential. Plus, if papers consider and address the use of oracles, an overview of the various implementation designs envisioned will be provided. Depending on the data gathered and the availability of implementation proposals published, Oracle's implementation will also be differentiated according to the specific accounting and reporting purpose. If a specific application is more advanced and shows higher robustness against the Oracle problem compared to others, it will be highlighted in this research.

The research questions of this paper are the following.

1) Is academic literature on blockchain and accounting integration considering the role of oracles and their limitations?
2) What portion of related literature is affected by biases and misconceptions about blockchain potential also for not considering oracle's role?
3) What types of oracles are proposed in the accounting field?
4) Which blockchain-based accounting integration shows more robust or advanced research on oracles' role?

Observing the results of this study we may argue that although this specific sector was not neglected by academic research, the scarcity of specific studies on the oracle's role raises dramatic concerns about the usability of the results obtained so far. Reporting for esg is the only sector in which oracles role and design has been considered and addressed, while also some interesting workarounds were suggested to bypass oracles use completely. Other literature section, such as adoption instead, does not consider at all the role of oracles. The remainder of this paper is as follows. Section two introduces the literature background providing an overview on past works on accounting and blockchain integrations. Also includes a specific section on most common misconceptions concerning blockchain technology which may help to better understand integration barriers. Section three describes the methodology and the variables chosen for data extractiona along their motivation. Section four, displays the quantitative results, discussing observable insights and section five outline the qualitative results. Section six discusses the overall content and elaborates common oracle schemes in the accounting field. Section seven concludes the paper, offering final thoughts on the study contributions and limitations, while providing insights for further research.

## Literature Background

### Blockchain technology

Blockchain is a generic term used to describe specific databases whose data is stored in sequential batches called blocks. One of the most known is the Bitcoin network, which is a unique blockchain type characterized by being open, transparent, decentralized, pseudonymous, and immutable to date [17]. Its security, decentralization, and immutability are guaranteed by a very strict consensus mechanism known as proof-of-work, which is based on a competition among actors called miners who use computing power to solve a cryptographic puzzle required to add new data blocks [18]. As the cost to generate the required computing power is considerably high and the success of block addition is not guaranteed, the chance for a malevolent actor to randomly manipulate the blockchain with false data is considered negligible. Plus, as the convenience of mining blocks strictly depends on exogenous factors such as the bitcoin price, cost of electricity, hardware availability/obsolescence, and government regulation, the composition of miners is constantly changing, guaranteeing a high level of decentralization [19].

As the term blockchain obtained notoriety thanks to the advent of the Bitcoin network, which, as stated above, is unique, it created the false belief that all blockchain wields the same or similar characteristics. If it's true that the Bitcoin blockchain is open, transparent, immutable, and decentralized, it is also true that a regular blockchain does not necessarily wield any of these [20]. Although counterintuitive, it Is important to specify that the decentralization of Blockchain does not depend on the fact that there are multiple nodes that store data, but there are multiple entities with the power to “add data.” Those who have nodes, in fact, maintain copies of the database on a voluntary basis and do not really contribute to its security, and for this, they are not rewarded. The security is guaranteed by miners or stakers who invest in mining equipment or stakes and compete with each other in exchange for some cryptocurrencies. It’s the harsh competition that makes the manipulation of database highly improbable, not the distribution of database copies [17]. Therefore, when comparing chains, it is important to understand how the power to alter database is distributed. The “blockchain trilemma” is a common concept used to explain these mechanisms, and is extensively explained in the following references [21], [22], [23], [24]

It is also important to note that as a ledger, Bitcoin keeps track of asset ownership by updating information about their property; it does not directly follow the assets to check who owns them. Wallets and coins on Bitcoin are only theoretical constructs for the system to be understandable by a broader audience. Figure 1 shows, in fact, the raw content of a block, which is just constituted by a data chunk from which no coins or wallets can be observed. Software known as Explorers is used to “standardize” the data chunks contained into blocks and generate a visualization similar to our legacy double-entry accounting to make it easily readable by a broad audience. Therefore, as Bitcoin is not really used to “trace” its assets and external software is used to enable this function, using it to trace and follow real-world moving objects could be a questionable choice [22][25], [26], [27]. A hilarious talk from a well-known Bitcoin popularizer, discuss the debatable choice of putting bananas on the blockchain [28].

Figure 1. Content of Bitcoin block n°840946

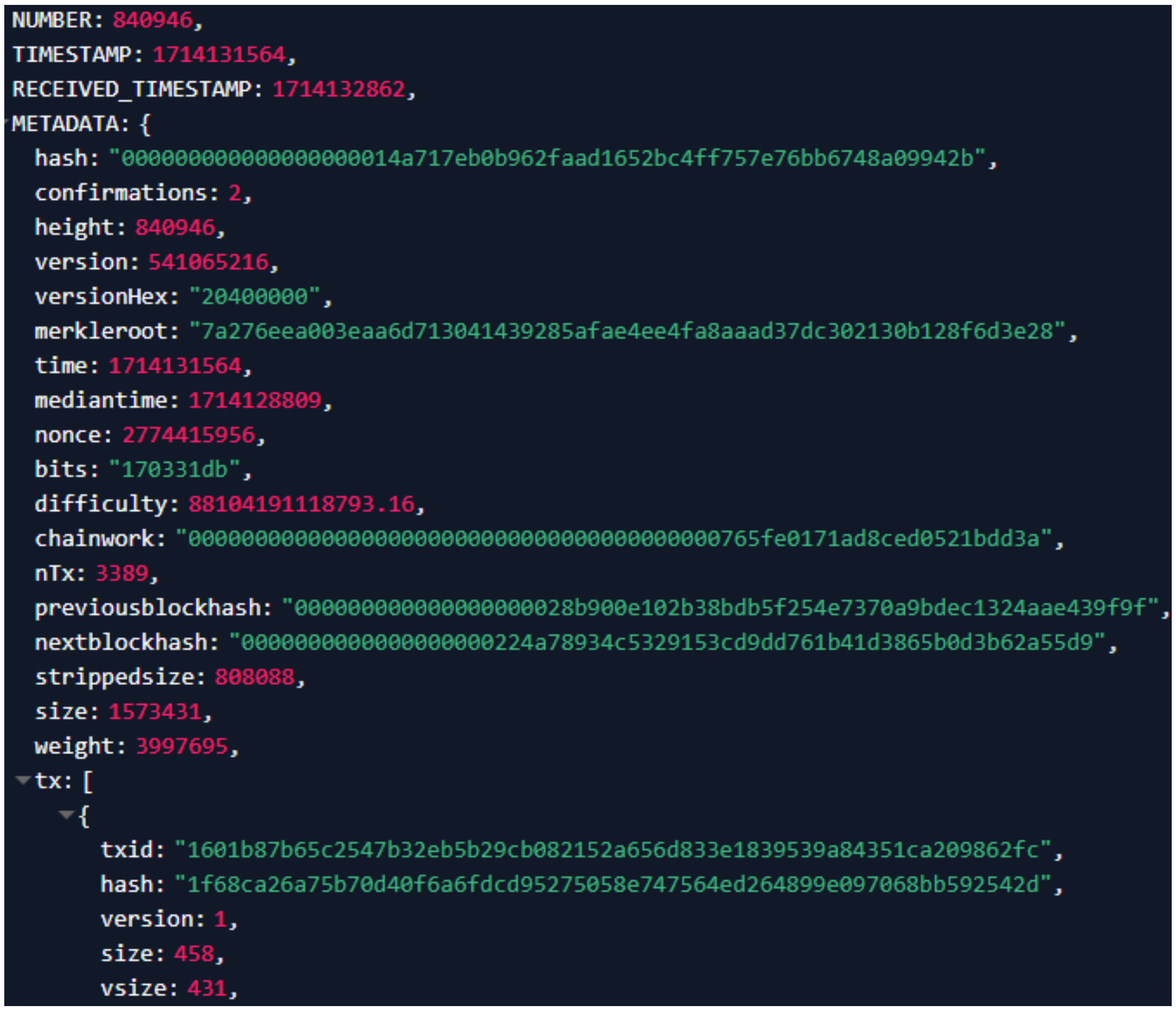

As blockchains are closed ecosystems, they cannot be directly implemented for real-world applications such as asset traceability, Intellectual property rights, academic and land records, digital identity, or legacy accounting systems [2], [29], [30]. The workaround introduced to allow blockchain to communicate with the real world is constituted by oracles, which are software or hardware that cover this specific purpose [31]. It is fundamental to specify that data introduced by oracles, as any other arbitrary data inserted on the chain, is not verified in its veracity by any nodes or consensus mechanism, which only verifies its input condition (e.g., the related fees have been spent). Therefore, it's trivial to understand that leveraging blockchain for a real-world application does not affect in any way the reliability or veracity of the data stored [15]. There are, however, some specific oracle mechanisms used for on-chain finance, insurance services, or prediction markets, whose function is to ensure that the data stored comes from reliable sources. Their design and development are, however, still in the early stages, and their failure has led to a dramatic loss of capital invested in related protocols [32], [33], [34].

Oracles on blockchain were introduced by Mike Hearn, a cryptographer who had the idea to leverage Bitcoin in real-world applications and needed a system to allow smart contracts to fetch data from the real world [35]. It is important also to take into consideration that back in the early days, names were given to these features almost randomly, and the name “smart contract” does not imply any smart/automation capability or legal value to these pieces of software [36]. A smart contract is, in fact, just a computer program that runs on a blockchain.

and as any third-party program they may contain bugs or misbehave [37][38]. Therefore if used on a blockchain, they negatively impact its underlying security and reliability [39]. To be secure and reliable, blockchain applications should have few or no smart contracts. The most secure blockchain (Bitcoin) has, in fact, very limited support for smart contracts.

One last thing to consider that can dramatically affect the usability of blockchain integration is that Blockchain, although de facto a database, is not meant to store, in a systematic way, data that does not pertain to transactions. Although on Bitcoin, can be retrieved all sorts of data, from poetries, song text, pictures, documents, and so on [40], Bitcoin core developers have fiercely opposed the practice of uploading external data on the chain in the early days, which affected the development of third-party applications and led to harsh internal debates, resulting in either hard forks (BitcoinCash, BitcoinSV) or new blockchains with different rules such as Ethereum [36], [40], [41], [42]. Despite the new platforms allowing the input of arbitrary data on-chain, the operation had to be costly in order to balance the maintenance cost of the chain [43]. Therefore, although not a misconception as it's perfectly possible to use blockchain to store data, hypothesizing a real-world application whose main purpose is relying on a blockchain to store its data is highly improbable. Table 1 summarizes common blockchain characteristics. We deemed it necessary to briefly summarize these common blockchain features to better understand the complexity of integrating them with a real-world application such as accounting.

Table 1. Common blockchain characteristics/features and limitations*.

| Blockchain Feature | Reference |
|---|---|
| Blockchain decentralization is guaranteed by miners or stakers who compete with each other investing in mining equipment or stake in return for cryptocurrencies. Multiple nodes do not guarantee decentralization and security. | [17], [38], [44], [45] |
| The Bitcoin ledger keeps track of cryptocurrency owners by adding updated data to the ledger. Since wallets and coins are theoretical constructs, blockchain do not really trace them around the network. | [25], [26], [46] |
| To be implemented for applications above the simple exchange of currencies, blockchain leverages smart contracts, which are third-party developed computer programs whose security and reliability are not dependent on the security of the chain. | [37], [39], [47], [48] |
| For applications that require real-world data, oracles are introduced, whose security is also unrelated to the security of the chain. On the other hand, the difficuly in building secure oracles is often leveraged to hack blockchain-based apps. | [2], [32], [49], [50], [51] |
| Blockchains store a very limited quantity of data. On Bitcoin, a 32-byte hash was already considered excessive, and although other chains allow a higher quantity of data, the price to do so is considerably high. | [40], [42], [52] |

*Author elaboration.

**Blockchain in Accounting**

Works of literature support the view that blockchain can significantly enhance the reliability and accuracy of financial documentation [53]. The openness and immutability of records when transactions are digitized [54] is seen as an mean through which fraud can be minimized [55]. Dyball and Seethamraju [56], also, propose blockahin technology as an harmonization layer to facilitate auditor's work and improve audit quality, by providing a more reliable and clear transaction record. However, Alex et al. [51], even theoretically supporting these positive views of integration, point out a lack of practical case studies and real-world implementations.

Some groups of studies focus on specific areas of blockchain in relation to accounting, which may also result in new accounting methods. Previous works of literature [57][58][59][60], distinguish the areas in those listed below.

**Triple Entry Accounting.**

Introduced by Yuri Ijiri [61] as a theory to measure the "momentum" of money, triple-entry accounting (TEA) concept was brushed-up by Ian Grigg [12] with a different meaning. Grigg proposed to leverage cryptography to limit errors and fraud in accounting. The basic idea was that companies should not be the sole recorders of business transactions, but a third party should have kept records of both transactions. These cryptographically secured entries should have corresponded to both the debit and credit of the two entities. Although revolutionary, by the time the concept was proposed, it was unclear how the third entry should have been recorded and managed and by which authority. The emergence of Bitcoin and blockchain technology in general suggested the idea that probably a central authority managing the ledger may not be necessary if a third-party ledger, secure, transparent, and immutable, is already available. Some researchers and also practitioners then started exploring this avenue with some early insights. Weiyi Cai [62] provided a review of case studies for projects proposing blockchain-based triple-entry accounting; the research shows that although real-world applications of TEA are scarce, the results obtained are already promising. However, important challenges have to be faced, such as the uncertainty of regulations and return on investment [63].

**Real-Time Accounting/Auditing and continuous auditing.**

Byström [64] argues that If market pressure made firms put all their business transaction on a blockchain with a permanent time-stamp on each transaction, the firm ledger would be instantly and openly available. This would allow anyone to aggregate the firms' transactions in real-time, creating the related income statements and balance sheets. Therefore, if firms keep all their transactions and balances on a blockchain, auditor's work may be facilitated and enhanced. Considering the fact that auditors must perform intensive work of preparing and harmonizing the accounting data before performing the audit, if the data has been all uploaded on the blockchain, this preemptive work is indeed easened [65]. The reconciliation work is practically done, thus reducing the chance for human mistakes [8]. Constantly uploading all data on the chain is sought to turn conventional auditing based on retrospective analysis into continuous auditing based on real-time data [66]. Other authors, also supports the view that the real-time availability of a large quantity of data may also enable AI-based predictive systems to detect anomalies and prevent errors [57]. There is also a shared vision that the real-time upload of accounting data will reduce the chance of manipulation and fraud ex-ante if managers knows that all transactions are immediately available, transparent, and immutable, as there will be no chance to alter them at a later time [67], [68]. O'Leary [69], however, argues that these integrations will work only if companies decide to record all the transactions on the blockchain. If only part of the transactions are recorded, the advantage is minimal.

**Governance, Trust, and Accountability**

Schmitz & Leoni [8] argue that blockchain has an impact on governance since in case accounting data is put on the blockchain, then stakeholders would have immediate access to it. When a considerable amount of accounting data over time has been collected on the blockchain, stakeholders can also easily inspect historical accounting data without cumbersome procedures. The immutability property of blockchain would also guarantee that retrieved data was never manipulated [70]. A digital signature can also help find the trail of data upload and the actors that performed the process, improving accountability and transparency toward stakeholders [71]. Lastly, research from Massaro et al. [72] also explores the use of NFT deployed on the blockchain to enable new forms of governance. Holders of these NFTs are, in fact, allowed to take part in business decisions that can also affect a company's future.

**Esg Reporting**

Above financial reporting, scholars have also debated on the possibility of using blockchain for non-financial disclosure. Critical in this field is that sustainability data is acquired by reporting companies from different entities along the supply chain whose reliability is inhomogeneous [73]. However, as described in the governance section, since digital signature is used to upload data on the blockchain, there will be at least the trail responsible for the data upload [58]. The optimal method of uploading non-financial data to the blockchain and the specific chain to be preferred for this specific data is still to be identified.

**Blockchain adoption and impact on accounting professions**

Other than exploring the consequences of integrating blockchain in accounting, the literature also investigated the direct effect that the advent of blockchain had on accounting and auditing professions. The acquisition of digital assets, such as cryptocurrencies, non-fungible tokens, and inscriptions, for detention, trading, or use in related metaverses within the business activity poses a real issue of their correct registration in the accounting books and consequently, their auditing [74], [75]. This clearly impacted regulations and required a new set of skills for accountants as well as for auditors to be prepared for these emerging use cases. There is no proof that the advent of blockchain will make accountants and auditors obsolete, but there is evidence of increased complexity for these roles [8]. Those that on the other hand support the idea of a reduction of importance for these professional roles, sees in blockchain the power to overcome reconciliation and guarantee the authenticity of transactions, de facto dramatically reducing the need for accountants and auditors. [1], [69], [76]. In light of these advantages, other research has investigated barriers and facilitators to firms' adoption of blockchain for accounting practices. [77], [78], [79]. However, this research focuses on the impact of blockchain on legacy accounting. Therefore, the impact of digital asset registration on the accounting and auditing profession, although a critical matter, is out of the scope of this study.

**Methodology and data extraction.**

In order to answer our research questions, building on [16], a systematic literature review is considered necessary and also sufficient for the intended scope, although further analysis is performed on the retrieved data. Systematic reviews are critical in synthesizing existing research, effectively identifying gaps in literature and establishing the state of the art within a field. Their structured methodology ensures a comprehensive and replicable analysis, which enhances reliability and supports evidence-based conclusions [80], [81].

The research started on 18 September 2024, and Scopus was identified as a database for this study. The following research string was used:

(TITLE-ABS-KEY(blockchain) AND TITLE-ABS-KEY(accounting) OR TITLE-ABS-KEY(reporting) OR TITLE-ABS-KEY(ESG) OR TITLE-ABS-KEY(auditing) OR TITLE-ABS-KEY(accountability))

The query returned 3175 results, from which 42 were removed for being non-English, 48 were titles of conferences, 72 were books, and 18 were notes. Retrieved articles were then inspected in the title and abstract to check pertinence to the topic under analysis. The core idea is to retrieve articles that involve the use of blockchain for legacy accounting auditing and reporting purposes. Therefore, accounting and auditing for cryptocurrencies or blockchains for bugs, for example, is clearly unrelated. After a first skim of the literature, 2776 were removed for being unrelated, which led to a sample of 219 articles. Unfortunately, seven articles were unretrievable, which forced a reduction of the sample to 212 articles. All 212 were then downloaded and read to retrieve the required data for the study. After carefully inspecting the articles, we realized that 30 were unrelated despite a first title abstract matching. This led to a further reduction of the sample to 182 articles, all of which were related to the topic under investigation. Above the data extracted typically for a systematic review, such as article type, date, paper type, topic, and so on, other categories are identified to meet the purpose of this study. Further building on [16], articles were inspected for the literature provided on blockchain and divided among them that considered blockchain as a univocal technology, or distinguished among public and private, for example, or given a deep distinction considering also different chains. The rationale is that since blockchain is not univocal; sharing this idea provides a biased expectation of technology potential. For the same reason, in empirical papers are also inspected if the related blockchain under investigation is explicated. The reason is that, the related findings are most likely not extendable to other chains.

The primary focus of this research, however, is to inspect the consideration of this piece of research for oracles and the oracle problem. Therefore, papers are also inspected for including reference to these subjects. Using just a word processor and inspecting for the presence of the word oracle or using the related database function to spot articles with the word in the text was not considered a viable option. As the taxonomy of oracles pertains mainly to the field of computer science, it's expected that oracles are also considered in research in an indirect way by mentioning sensors, RFID, or nodes. Therefore, the paper's text is scrutinized for direct or indirect references to oracles. Plus, the presence of the word oracle could also be due to a homonym, for example, of a renowned database company, or simply for a citation or reference. Therefore a manual inspection was the only possible alternative. As for the oracle problem, the approach is the same. Plus, also in computer science, can be retrieved instances of reference to the oracle problem as "garbage in, garbage out" or oracle paradox, therefore the text was inspected to spot any reference to the unreliability of oracles or related issues. Articles were then differentiated to refer directly or indirectly to oracles and the oracle problem. In [16], there is also a speculation on the fact that articles mentioning and including oracle and oracle problems are less likely to include misconceptions; therefore also, this aspect is investigated. Misconceptions may include the idea of blockchain ensuring the truthfulness of data, automation capabilities, or real-world integrations without considering the role or limits of oracles. A recent article by Carol Sargent [82], for example, shows that many papers published in the accounting field include misconceptions about blockchain consensus. Therefore, it is important to investigate further the nature of these misconceptions and whether there are other types of biases. On this specific aspect, the author wishes to state that the presence of misconceptions, although clearly showing a theoretical limitation of the current research, does not in any way judge the quality of author's work or good faith. The provenance of biased data may also come from review, so articles may rewrite information found in other articles in good faith. Interviewed experts may also provide misleading information that is reported by authors in good faith. Since blockchain is an extremely hard and complex subject to investigate on which official and reliable information was hard to find, especially in the early days, a considerable presence of misconception is understandable but clearly not desirable. Again, as suggested in [16], the data extracted from the literature review will then be analyzed in relation to the consideration of oracles used as a pivot. Figure 2, displays the research design.

Figure 2. Research Design.

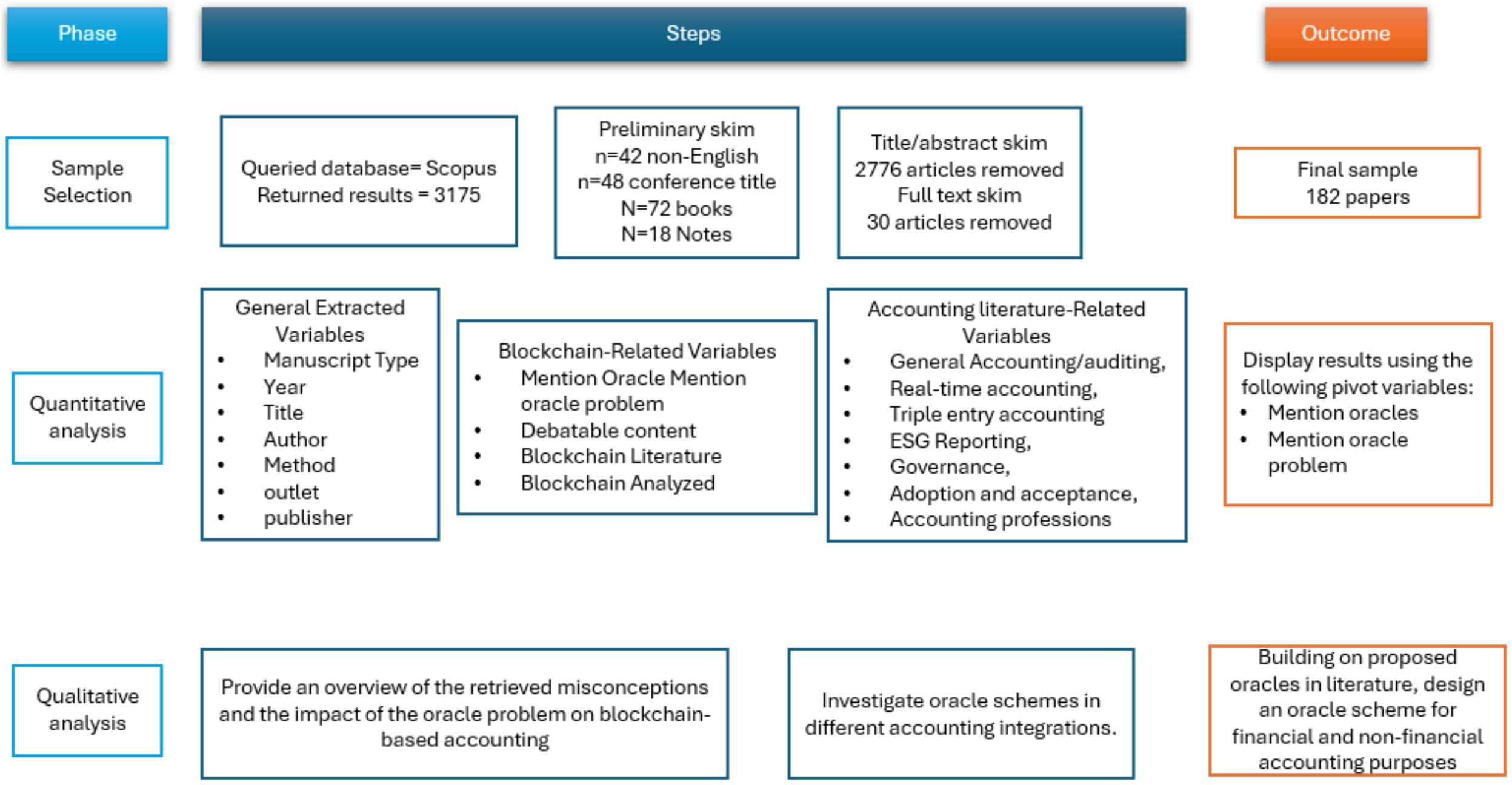

## Findings

Although the core of the analysis revolves toward the consideration of oracles, the overall data already provides some insightful suggestions on the state of the art. Considering the global sample, we can already observe a dramatic scarcity of empirical research in the field before adding the oracle filter. Of 182 papers, 57 are theoretical, 37 are reviews, 18 are proposals but also theoretical, and the other 70 papers involve some primary data. However, of these 70 research papers, 38 use qualitative data from interviews, so the actual experiments, case study, and research on accounting data with 33 entries count for less than 18% of the sample. If we consider the time frame to be recent but not that narrow, 33 practical studies of which only 5 are case studies show a dramatic lack of academic research of an empirical nature performed in this field. Another interesting information that emerges from the whole sample is the degree of generalization of the blockchain concept. As explained, since blockchain is heterogeneous, general research on blockchain is hardly meaningful if specific constraints are not extensively explained. Data shows that 51% of the sample sees blockchain as univocal, while 18% distinguishes between public/private and consortium, and only 30% provides an extensive distinction among different blockchain types. Considering research that specified the blockchain investigation, the percentage is alarming. Apart from 83 papers made of reviews or research on accounting professions in which focusing on a specific chain could be considered not of primary importance, other 66 papers made of proposals and case studies do not specify which chain they are investigating, therefore making their results and data hardly replicable or inferable for further studies. The few (33) papers that focused on specific blockchains display a slight preference for Hyperledger (13) and in general private (4) or consortium chains (2), while another group focuses on public blockchains with Ethereum (7) as preferred chain followed by Bitcoin (2) and BitcoinSV (1). Concerning specific use, literature shows preference for Ethereum public blockchain for general accounting and auditing purposes, while a permissioned chain such as Hyperledger is preferred mainly for ESG reporting purposes.

The last interesting data that can be observed from the whole sample is the presence of articles containing misconceptions. More than half of the sample (74%) does not contain any of these, while 26% include several misconceptions. Less than one-third of the studies published on this subject include conceptual flaws, which is actually not negligible information. As expected, most of these misconceptions concern the fact that, for some reason, blockchain affects the veracity of data.

Beyond this already insightful information, even more intriguing results emerge when focusing on oracles and the oracle problem, using them as a pivot to analyze the remaining data. First data shows that in 123 papers, 67% of the sample, oracles are not mentioned or considered, while 43 consider them indirectly and only 16 consider them directly. More alarming is the consideration on the oracle problem, which sees 133 (73%) articles not mentioning the issue, 40 mentioning it indirectly, and only 4 discussing the problem directly. As a matter of fact, mentioning or stating the problem directly or indirectly does not affect the quality or the meaningfulness of the research delivery, as its just a matter of taxonomical choice. Furthermore, oracles is a term used mainly in computer science, therefore is for research in accounting is understandable that other synonyms such as sensors or IoT are used. The situation is more evident for the oracle problem, as in computer science, other synonyms such as oracle paradox and garbage in garbage out are used. The real, critical data in this case is the absence of these concepts (directly or indirectly) in nearly 2/3 of the sample.

Considering oracles and the oracle problem as a pivot, the other data shows interesting results. Date and outlet as expected do not really affect the inclusion of these subject as their appearance is constant across the timeframe. An interesting distinction instead emerges from the methodology type in which proposals and case studies mostly consider the role of oracles and oracle problem. The practical nature of these studies most probably obliges to provide insights on the communication channel used to transfer data into blocks. On the other hand, empirical research using interviews and accounting data have almost no consideration for the role of oracles, showing an important literature gap. Theoretical papers also rarely consider the role of oracles, and probably due to the general lack of consideration, reviews rarely mention oracles and their role in the accounting field. The literature background also influences the consideration of oracles, as papers with univocal descriptions of blockchain rarely consider oracles, while their role is mostly addressed in papers with extensive descriptions of different blockchain types. The distinction for blockchains also shows an interesting insight. Papers that include the analysis of a specific blockchain type, regardless of which one, are most likely (68%) to include a section discussing the oracle's role.

A very interesting insight is obtained considering the presence of misconceptions in combination of oracle's consideration. The chart shows that the percentage of articles containing misconceptions drops significantly, from 25% of the whole sample to 6% of the sample of papers considering oracles and the oracle problem. It is, therefore, arguable that papers considering the role of oracles and their limitations are less likely to be theoretically flawed by common misconceptions.

The topic analysis also shows a worrying situation. A part of papers on general accounting and auditing that for being consistent in number count also some contributions considering oracles and the oracle problem; other specific applications count no more than one contribution on average. The only sector that counts more contributions is the ESG reporting with 18, followed by accounting professions with 8 and triple entry accounting with 7. Although it can be debatable if the knowledge of oracle can affect or not the adoption and acceptance of blockchain, which should, however, be indeed part of the analysis, it should raise no doubt that their role is of key importance in applications concerning governance and accountability purposes.

The last piece of information that is undoubtedly worth attention is the distribution by publishers. While Elsevier is the publisher with most publications considering oracles and the oracle problem, followed by IEEE and AAA, none of the second-tier publishers have papers that consider these subjects. On the other hand, 60% of publications containing misconceptions come from second-tier publishers. Although expectable, this is clearly a sign that the rigorous and strict reviewing and screening process and experience of reviewers of first-tier publishers inevitably affect the content and quality of published manuscripts.

Above the quantitative insights inspected in this section, the retrieved manuscripts offer qualitative data that is discussed in the next paragraphs.

**The oracle problem in accounting and auditing.**

Before delving into oracle designs in accounting, an overview will be offered on the ascertained oracle problem within blockchain-based accounting integrations explored in previous research and related misconceptions emerged. Many authors have warned about the issues arising from this integration of blockchain within the broader accounting and auditing sector. Gauthier and Brender [83], explain that "it is not appropriate for auditors to assume that...information recorded in a blockchain can be relied on with no prior testing". They further warn that the auditing standards take time to change and tend to change only when incidents happen that challenge the auditors' work.

An interesting research by Autore et al. [10], in fact shows that due to the false beliefs in improved reliability given by blockchain technology, earning management is facilitated and more likely to happen in companies that integrated this technology. The concept of truth on blockchain for those who have a limited background in computer science is clearly counterintuitive. All the data on the blockchain is indeed "TRUE," but this means that it has been added in compliance with specific conditions. Information evaluated as "TRUE" is not necessarily accurate or reliable. Nothing prevents writing falsities on the blockchain as long as the condition to write this information is verified [15]. The Bitcoin blockchain was initially meant to prevent double-spending and provide certainty of time for transactions, but no references are available on the intention of using this tool to provide veracity of information, nor does it support this specific purpose. The conviction that blockchain technology influences the trustfulness of accounting data, is indeed an important issue as this misconception is arguably broadly spread in literature.

Goel and Mishra [9] consider, in fact, blockchain as a “canonical source of truth.” Yang [84] confirms that information on chain is “authentic and cannot be changed at will”. Zheng [85], and Matringe and Power [86] states that blockchain encryption mechanism ensures the reliability and authenticity of accounting information: “algorithm as truth of fact”. Remarkably, even a study involving external experts includes misconceptions about data truthfulness, according to an interviewed expert who declared, “I believe that transactions recorded would be mostly authentic”[87]. This confirms the concerns already raised in previous research [22] that the quality of experts in these studies needs to be verified more accurately.

Concerning the reason why there is a misconception about the truthfulness of data on the blockchain, Carol Sargent [82], argues that: “*the potential disconnect between digital and real-world transactions can create potentially misleading illusions-of-truth*”. Basically, the article explains that consensus does not mean verification and shows that 58% of the inspected articles rely on this misconception. Extending the critique of Sergent [82], Coyne and McMickle [88] explains that “*The maintainers of these blockchains know nothing about the true validity of the transaction....the agreement between the two parties that resulted in the asset transfer. They only know whether the transaction uses unspent inputs and is digitally signed*”. A consistent stream of literature, however, supports the truthfulness of accounting data on the blockchain relying on the consensus mechanism. Liu [89] states that if accountants want to commit fraud, they must record false data on the blockchain, broadcast this to all nodes, and obtain consensus, saying that this is nearly impossible. Zhong and Fan [90] add “if a node has false information, all the information reported on this node will be rejected by other nodes”. Although this concept is also supported in [91][79][92][93], nodes do not verify the truthfulness of data uploaded data uploaded by other nodes. Therefore, as explained in Lobanchykova et al. [94], although, to a certain extent, we can exclude manipulation ex post [95], we cannot be certain of the historical truth of accounting records. Another stream of research proposes accounting integration relying on smart contracts and automation capabilities of blockchain, which, unfortunately, doesn’t possess. As specified by Liu et al. [3], implementing blockchain in auditing through smart contracts will increment the work for auditors, who also have to verify the reliability of these contracts. Authors in [96][53][97][98] envision in blockchain and smart contracts the ability to monitor transactions and predict the eventuality of mistakes or fraud. Although possible with some oracles connected with AI tools digesting blockchain data it cannot guarantee any transparency or fraud-free mechanism. Again, the automation of accounting action on the basis of predetermined rules proposed in [99] is not doable with blockchain and smart contracts alone since they are not connected to real-world data. Described in [100] as a first and last-mile problem, the author argues that blockchain cannot verify the truthfulness of data acquired through a third-party, and similarly, an auditor cannot be certain of data retrieved on the blockchain [101]. Also, the “track-and-trace” blockchain capability advertised in [6][102][77] that should be leveraged in continuous auditing is a false myth that cannot, unfortunately, be leveraged for this purpose.
Further studies support optimized auditing function with blockchain thanks to the reduction of intermediaries, which should alleviate or eliminate the work of auditors [103][104][105][106]. Tan et al. [107], however argues that "is inconclusive that using a blockchain-based AIS will automatically produce financial statements that are true and fair, and hence the fear that the audit industry will be disrupted is unfounded". As extensively explained in previous research [108], [109], due to the blockchain reliance on oracles for external data, we do not attain a reduction but a substitution of intermediaries. Despite the mistrust in some financial institutions, there is no reason to believe that third-party blockchain oracles are more trustworthy and reliable intermediaries. This will de facto not alleviate the work of auditors which will have to verify the reliability of third-party oracles and the genuineness of transmitted data. Finally, a last stream of research bases its integration idea on the possibility of registering all accounting transactions on the chain [85][64]. As explained, although possible in theory, this is undesirable. Even neglecting the high cost in fees, if blockchain is thought of as a "commons", freely available to the public, but in which the space is limited, registering accounting data on chain, will dramatically limit the usability of blockchain for other use cases, plus obliging nodes to maintain these accounting data at their own cost. Therefore, as has already happened in the past for non-standard transactions, this is unlikely to be allowed [40], [110]. Due to all these false beliefs and limitations, Coyne [88] argues that blockchain cannot serve an accounting purpose. We do not share the same pessimistic vision. The problem is integrating blockchain in the proper way, for the right purpose, and with the right oracle. The next paragraphs explore what literature proposed as oracle schemes for blockchain-based accounting purposes. Table 2 provides a breakdown of misconception effects in the blockchain-based accounting field.

Table. 2 Misconceptions and oracle problem in blockchain-based accounting.

| Misconceptions in blockchain-based accounting. | References | Clarification | Reference |
|---|---|---|---|
| blockchain encryption mechanism ensures the reliability and authenticity of accounting information | [86][84] [85] | Blockchain guarantees immutability up to a certain extent, not truthfulness. Oracles can verify truthfulness only if a specific design is implemented | [83][10][94][111] |
| if a node has false information, all the information reported on this node will be rejected by other nodes | [90][91][79][92][93] | Nodes are not aware of the true validity of the transaction. They verify if the input is unspent. Arbitrary data inserted by oracles is not verified for validity. | [82][88][30][49] |
| Smart contracts can automatically monitor transactions and predict fraud. | [96][53][97][98] | Smart contracts cannot provide more automation than any other regular computer program. Oracles can perform off-chain computation but are not infallible. | [29], [38], [39][47][112] |
| Blockchain can reduce intermediaries and eliminate the work of auditors. | [103][104][105][106] | Auditors are essential to evaluate the reliability of third-party oracles and cant be eliminated. | [3][101][107][113], [114] |
| Blockchains can be leveraged as a decentralized database for accounting transactions | [85][64] | Blockchains have extremely limited data storage. Storing large quantities of data can be expensive, as it is opposed. | [40], [110] [40], [41] |

**Oracles in Accounting and Auditing.**

Studies investigating in practical terms blockchain integration in accounting are scarce in number. This makes the availability of paper describing oracle schemes even narrower. As for Triple-Entry-Accounting, two examples of possible oracle implementations come from Craig Wright, a controversial figure known for claiming to be the creator of Bitcoin. In Pan et al.[115], the authors propose linking wallets with identity and registering invoices on Bitcoin so that auditors can easily verify the correctness of invoices and their provenance. Sunde and Wright [116], extend this concept, and propose a "fingerprint" with a digital signature that has to be associated with every double-entry transaction and copied into a third ledger. The third ledger will then have both the fingerprint of the related parties' books and their signature for auditing purposes. In the author's words: "*This way, identity can be verified without relying on a single oracle or intermediary*". Cao and Tsay [11] propose a system of triple-entry accounting in which both parties involved in a transaction have to sign a smart contract that is then verified by a "miner" who stores the transaction on the chain for auditors' inspection. In this context, the certainty of the transactions is guaranteed by both enders and receiver signatures, while the timestamp does not allow earning management. Buyers and sellers are themselves oracles. Other research on Triple-Entry-Accounting, such as Faccia et al. [117], [118] and Maiti et al. [119], although mentioning a mechanism to put data on chain manually or with APIs, do not further elaborate on that, and therefore, no speculation is possible on how it can be replicated in other systems.

Rozario and Thomas in a well-cited article [120] also discuss blockchain's potential to enable continuous audits. The basic idea is to use blockchain as an immutable ledger, with the advantage of having a traceable origin of data, improving accountability. Furthermore, with IoT connected to blockchain, much more non-financial information can be made available for auditors for a wide range of purposes. Barandi et al. [121], further specifies that information and resources should be shared directly between peers, rather than through a central node. McCalling et al. [122], extend this idea for continuous auditing but particularly focus on data certainty. The proposal relies on multiparty security in which data is scattered among customers and is aggregated with this technique on the blockchain, with the aim, in the author's words, of "emulating a trusted party". The communication channel, however, still represents an unsolved challenge.

Concerning specific applications in accounting, Goharshady [123], focuses research on credit reporting and elaborates on identity management to allow access to data on the blockchain. The idea is that users will create a record with their private key and a "fingerprint" that constitutes a unique identifier, and the lending institutions will add credit risk data. Other institutions will then need keys from the user (not the lending company) to access the record. Therefore, in this case, the lending institution performs the oracle role that registers the user's credit risk data. Sarwar et al. [124] propose instead data vaults for trial balance.

The idea is to hash each line of trial balance and store it in a cryptographically secure place to verify if the trial balance registered in AIS is then manipulated. The ownership and responsibility over data remain in the hands of the company. Therefore, the oracle is the company itself. More focused on tax, Lazarov et al. [125] propose a system for tax reporting in cross-border transactions. For this system to work, both nationalities involved in transactions should sign a copy of the same transaction with different identifiers. That way, for each transaction, there will be two copies, one with a country identifier and another with another country identifier. If one transaction misses its hashed copy, either there was a manipulation or a misreporting. Again, oracles are the authorities themselves.

More oriented to a general integration of blockchain in accounting, Wu and Cheng Li [126], proposes a system of smart contracts to link economic events, accounting events, and accounting reports. The proposed method to gather real-world data is a system of IoT and Gps that tracks goods from buyers' and sellers' warehouses. No additional explanation is given of how they should work, and the authors state that "*The IoT serves as a data collector and recorder immune to mistakes and fraud, which ensures the faithful and prompt upload of inventory information to BC*".

Mingming [127] also describes his blockchain integration idea with an I/O interface that transfers data on-chain, which is securely stored to prevent manipulation. Again, no more info is provided on the I/O Interface. Anwar et al. [128], bypass this issue, assuming transactions happening on the blockchain directly and suggesting nodes have the company name so that the public key can be recognizable for the auditing procedure. Ingle et al. [129] also, propose a network of companies of which any represents a node, and information on the chain is inserted upon their consensus, although not explaining the consensus type.

The only resource that discusses oracle within the accounting profession is from Sheldon [113], [114]. The author does not propose a specific oracle scheme but envisions the auditors as oracles themselves. The author argues that oracles should be treated as service organizations under auditing standards because they perform critical data processing tasks that affect financial reporting and control systems of the entities using them and propose control objectives for auditors to ensure the integrity of data used by smart contracts. Open questions and challenges for accounting professionals are also identified. These include defining what constitutes an appropriate consensus among multiple oracles and how to handle the risks of compromised data sources. The author then warmly invites auditing professionals to understand and address the risks associated with oracles, as firms using blockchain must ensure that oracles providing data to smart contracts are reliable and have appropriate controls in place. Despite the brilliant contribution of Sheldon in 2021, none of the following articles concerning the adoption of blockchain in accounting or its impact on the accounting profession include references to this crucial aspect. Table 3 summarizes the proposed oracle design in accounting and auditing.

Table 3 Proposed oracles in accounting and auditing.

| Oracle Mechanism | Accounting Application | Reference |
|---|---|---|
| Buyers and sellers sign the transaction and take responsibility for the veracity of data. | Triple entry accounting | [11][115] |
| A fingerprint is associated with the transactions in the double-entry ledger, and then both fingerprints are inserted in the third-entry ledger. | Triple-entry accounting, general accounting, and auditing | [116][123] |
| Use IoTs to put data on chain. | General accounting and auditing, Real-time accounting | [117], [118][119][126][127] |
| An authority signs the transaction and takes responsibility for its veracity | General Accounting and Auditing, credit risk reporting, tax reporting | [120][123][124][128] |
| Scatter data among peers and use multiparty security to mimic a trusted party. | Real-time accounting | [122] |
| Auditors are oracles themselves. | Accounting professions | [113], [114] |

**Oracles in ESG reporting**

Although there are studies [130] that envision in the simple use of blockchain higher ESG data reliability, there are others that actually investigate the issue of how to transfer this data on-chain securely. A stream of research conducted by the same investigation team, recognizes the persistent challenge of guaranteeing that the data collected prior to the upload on chain is authentic, debating the fact that little research is done in this direction [131][132][133]. Wu et al. [133], similarly to Liu et al. [134] propose a system of token-based incentives to motivate companies to increase ESG disclosure. In this specific study, they propose a combination of three variables to decide the amount of tokens and the value of the distribution among the companies that increase the disclosure. The data is evaluated for its interval of disclosure, throughput, and authenticity. The interval and throughput are capped as an excess in these two variables, which is not desirable from two points of view. First, an excess of data will result in redundancy that will reduce its utility; second, it will impractically congest the blockchain for data of uncertain utility. Therefore, the variable that has more weight is authenticity, whose evaluation and study are performed in further research. To support data authenticity, Wu et al. [132], propose a system organized in multiple layers to collect and transfer ESG data on-chain. The first layer, intended to measure the key performance indicators for ESG disclosure, as also proposed in [135][136] is made of meters and sensors to monitor electricity consumption, water consumption, gas emission, and waste. The second, interoperation layer serves as a support for data transmission. The data collected by the meters and verified in terms of space and time, is transmitted through network communication protocols such as Bluetooth, 4G/5G, Wifi, in batches, to the blockchain layer. The envisioned blockchain is a consortium chain, that should guarantee the inability of a single entity to manipulate the data and a certain degree of agreement in data upload. Plus, integrated APIs should allow the inward and outward transmission of collected data. The fourth layer is made by an interface that allows the data to be collected by intended stakeholders.

Chen et al. [131], further elaborates on ways to ensure the authenticity of data and proposes an event-based authenticity algorithm. The system is articulated in three phases of which, the first phase evaluates the responsiveness and reliability of IoTs, while the second phase collects data on different events distinguished by location, and the last phase analyzes the eventuality of correlation among events and data collected to further inspect the reliability of the data collected. The data is then re-evaluated by comparing it with a global index that is leveraged to spot possible fake or unreliable data.

Again, to guarantee the authenticity of data, but with a different approach, Heiss et al. [137] propose an attestation service for carbon emissions accounting and reporting on the blockchain. The article brilliantly explains that monitoring, reporting and, verification for carbon emission is unfeasible to do on-chain due to blockchain's privacy and scalability limitation that clashes with the privacy, quantity, and computation required for this type of data. However, blockchain characteristics can be leveraged if complex computation activities are performed off-chain. In their design, the system works as follows. In the monitoring phase, the carbon footprint data is obtained with attestation from the sensors and suppliers. Attestations serve to ensure that confidential inputs originate from the expected data source. In the reporting phase, the attestations are verified, and the emissions are calculated. A ZkProof is then generated and sent to the blockchain contract. Finally, the contract should inspect the ZKproof and evaluate if the product's carbon footprint has been correctly assessed in compliance with the accounting standards. Plus, in this last phase, it is verified if only data already checked by the auditors has been utilized for the parameters. Tang and Tang [138], along with Luo et al. [6], support instead a different approach to sustainability reporting. The underlying idea is to shift the database used from a centralized to a decentralized database to record voluntary disclosure of emissions. The idea is not to improve the security ex-ante but the security ex-post and also to allow ex-post easier data computation thanks to the accessibility of DLTs. In this sense, the responsibility for data authenticity has not shifted to the blockchain ecosystem but remains with the company responsible for the disclosure. Additionally, Luo et al. [6] propose that if authorized stakeholders are in charge of putting data on the chain, with their digital signature and time stamp, accountability in case of fraudulent data will be facilitated. In this sense, blockchain will not prevent false data upload but will help detect the accountable person and, up to a certain extent, help disincentivize malpractices. Table 4 provides an overview over oracle solution proposed for esg reporting

Table 4. Oracles for ESG reporting

| Oracle mechanism | Reference |
| --- | --- |
| System of IoTs, continuous data gatherings, and multiple checks with complex algorithms to ensure the reliability of IoT, eliminate outliers, and ensure data veracity. | [131][132][133] |
| An attestation service is used to guarantee that non-financial data comes from the designed source. | [137] |
| Companies or authorized stakeholders are themselves oracles and keep full responsibility for the data uploaded. | [138][6] |

**Discussion.**

Despite the numerous studies in the field of blockchain-based accounting integration, the number of articles that consider oracles and the oracle problem is exiguous. The literature is also plagued with misconceptions that are retrieved in almost two-thirds of the sample. The most common is the idea that blockchain positively affects the veracity of data, and as anticipated in Sargent [82], the shared reason is a misunderstanding of the consensus mechanism properties. The real mechanism to ensure data veracity, oracles, is on the other hand, mostly neglected, and although a decent portion of the paper mentions them, the qualitative analysis shows that just a fraction of articles provide further information on how these oracles should work. In the broader accounting and auditing setting, oracles are usually identified as authorities that directly upload the data on chain [138][6], or buyers and sellers themselves, who both sign the transactions [11][115]. The veracity of data is thought to be dependent on the fact that these agents upload their digital signature together with the data, and therefore, the chance for fraudulent data to be traced back to the responsible agent should serve as a deterrent for malevolent actions. The problem in this case, however, as extensively explained by Douceur [139] is to be able to univocally associate a digital signature to a real-world identity and that there is an effective accountability connected between them. If it's true that a piece of data can be traced back to a signature, it is not also true that this signature can be traced back to a real-world individual who takes responsibility for this action.

As for IoTs, research still shows confusion on the fact that these systems could be somehow more reliable if connected to the blockchain [131][126][127][117], [118][119] and provide trusted data. In the ESG reporting field, however, research is more advanced on this point of view as multiple works investigate the possibility of improving the reliability of these devices for non-financial information. A stream of research presents a complex design, divided into multiple phases to ensure the reliability of IoT devices and the truthfulness of reported ESG data [131][132][133]. Although this system appears to be robust in theory and indeed an interesting proposal, we argue that the complexity of design makes it highly costly and cumbersome for companies to implement, also considering the voluntary of non-financial disclosures. Therefore, despite the originality and value of this idea, it's unlikely to be implemented. Unfortunately, there are not many more alternatives proposed in the literature. One of these few comes from Heiss et al. [137] who proposes an attestation service leveraging ZKnowledge. The system ensures that data on chain comes from the designed data source. This is actually a valid Oracle mechanism. However, we should consider that there are services and systems already available and tested to perform the same activities. Provable Things is for example the first blockchain oracle mechanism that provide attestation service for data on chain, and also Chainlink can be leveraged for this purpose [140], [141]. Therefore, if the idea is to propose an attestation service and these are already available and working, it would be interesting from a research perspective to first test these mechanisms already available. Lastly to improve ESG reporting there is the proposal to put all the data on chain. As already mentioned, this is unlikely to be done on public chains given the limited storage capacity, but it is possible with private chains such as Hyperledger. Therefore considering all the oracles proposed in literature, and their limitations, an ideal ecosystem for either financial and non-financial reporting is constituted by a private chain such as Hyperledger, that leverages an attestation service such as Provable Thing or Chainlink, and in which the signature for data provenance is attached to a real-world authority that takes accountability for the truthfulness of data stored. The system is outlined in Figure 3.

Figure 3. Preferred oracle ecosystem for Blockchain-based accounting.

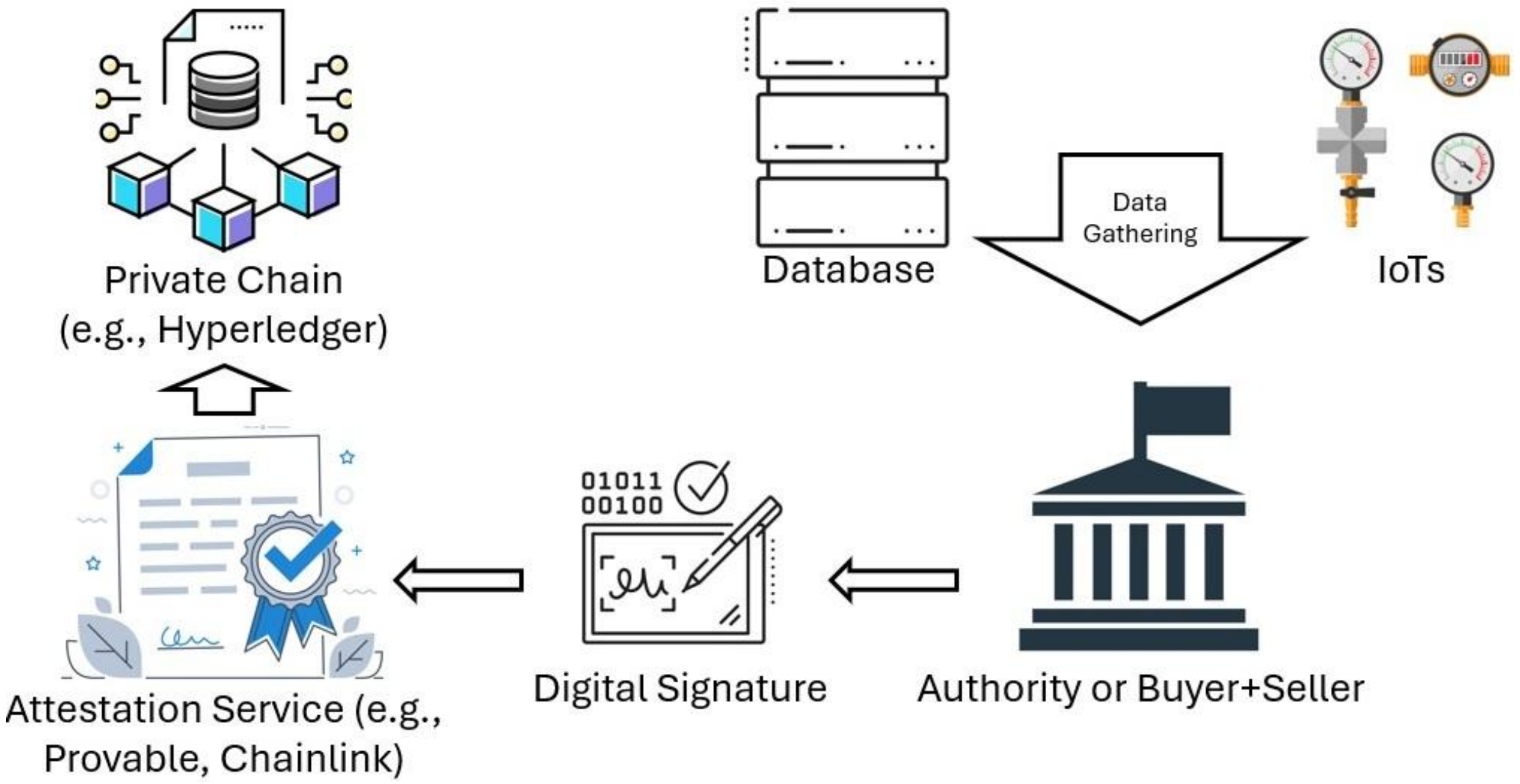

The proposal of McCalling [122] to leverage multiparty authority to ensure data authenticity is also valid. The only issue is that it assumes the existence of a "multiparty," which, although desirable in theory, is not always available in practice. In the case of non-financial disclosure, it is highly unrealistic to assume the availability of multiple data sources for the same piece of data. As for the limitations of the idea of putting all the data on chain, we would like to specify, however, that in the case of the adoption of central bank digital currencies (CBDC)[142], the transactions would happen directly on-chains controlled by institutions, and therefore, this will more easily and directly enable continuous auditing that is at the moment unfeasible with public chains such as Bitcoin.

An important gap in the literature concerns the availability of oracles schemes concerning blockchain integration in governance. Truthfully, there is a scarcity of empirical research in general in this field, and therefore, it was expectable the absence of more specific works, including oracle schemes. From what emerges from the literature, however, the effect on governance is more indirect and is expected from a possible disclosure of financial and non-financial information on-chain that should allow stakeholders to have more complete control over company data. Concerning instead the possible use of NFTs to participate in the governance of a company, the main issue would be to ensure that the owner of NFTs takes accountability for their choices, but this would be easily enabled with KYC. Anyway, given the infancy of the blockchain-based governance system proposes to date, oracles related analysis may be a bit untimely. On the other hand, what is highly worrying is the absence of oracle focus in research concerning blockchain acceptance and impact on accounting and auditing professionals. Excluding the brilliant contributions of Sheldon in 2021 [113], [114], the literature shows a total absence of further investigation on these topics. The issue is quite serious because if on the one hand, we may argue that due to the blockchain limitations discussed in this paper, the actual usefulness of integrating accounting on the blockchain is debatable, on the other hand, accounting for cryptoassets is an actual issue. Now, since the price of cryptoassets is given by oracles, if accountants or auditors ignore their existence or their use, or worse their possible manipulation, they will be unprepared for their proper accounting and auditing. Table 5 summarizes the key findings of this research divided by research questions.

Table 5 Key findings of this research.

| Research Question | Answer |
|---|---|
| Is academic literature on blockchain and accounting integration considering the role of oracles and their limitations? | 32% of papers in the final sample include content on oracles and only 17% mention limitations about oracles. |
| What portion of related literature is affected by biases and misconceptions about blockchain potential also for not considering oracle's role? | 25% of the whole sample includes misconceptions about blockchain potential but considering paper mentions oracle problem just 6% include misconceptions |
| What types of oracles are proposed in the accounting field? | Oracles role is expected to be covered either by:<br>• A designed authority/Stakeholder<br>• Buyer and seller themselves<br>• A combination of nodes through multi-party authentication.<br>• An attestation service.<br>• An IoT<br>• A system of IoTs, plus a system of algorithms to guarantee the authenticity of data. |
| Which blockchain-based accounting integration shows more robust or advanced research on oracles' role? | ESG reporting literature shows already advanced oracle mechanisms, although in very few papers. Practical implementation is feasible.<br>The study of oracles is, however, totally absent for |
| | accounting professions that should instead be prepared to audit for price oracles. |

**Conclusions.**

The present study extends the work of Carol Sargent [82], providing a more specific explanation of the limitations within the current literature concerning blockchain integration in the Accounting field. Above the cited misconception on consensus, the paper also distinguishes other common misconceptions and provides an overview of the most retrieved ones. Plus, building on [16], it inspects the current literature in accounting for consideration of oracles and the oracle problem, providing an overview of proposed oracle schemes and the most common design. The literature shows a rudimentary oracle design for blockchain integration in accounting that does not adequately support further studies and research. Some studies on ESG reporting, on the other hand, display advanced oracles schemes that could be actively leveraged to develop some proof of concepts. This study also shows the absence of Oracle consideration for accounting professions, which, given the urgency of accountants and auditors to be prepared to account for crypto assets, is very worrisome. In light of finding a workaround for missing oracle schemes, many studies take for granted that transactions happen directly on the blockchain, which is highly unlikely to happen in the short term and with the use of current cryptocurrencies. However with the advent of CBDC, transactions are not only digitized but also stored in real time in public ledger, therefore easily enabling the smart audit integration literature is proposing. An important limitation of this study is that it does only consider academic production and not practitioners work. In many academic papers, there is mention that big auditing companies have already integrated blockchain in this field, but there is no mention of the type of Oracle scheme implemented [74], [78], [87]. Further study should then focus on investigating how these companies integrated blockchain and with what Oracle design. Given the early stage of the few empirical papers in the sample, we may argue that research on this topic has barely started. The important decision is, however, if it is the case to continue investing in these research topics. Although it is technically possible to transfer accounting on blockchain, the benefits are still unclear and uncertain.

**References.**